\title{Exploring Genome Characteristics and Sequence Quality Without a Reference}
\author{
        Jared T. Simpson \\
        Ontario Institute for Cancer Research \\
        jared.simpson@oicr.on.ca
}
\date{}
\documentclass{article}

\usepackage{hyperref}
\usepackage{amsmath}
\usepackage{graphicx}
\usepackage{subfig}
\usepackage{cite}

\newcommand{\algo}[1] {
\texttt{#1}
}

\newcommand{\ms}[1] {
\texttt{#1}
}
\begin{document}
\maketitle

\abstract

The \emph{de novo} assembly of large, complex genomes is a significant challenge with currently available DNA sequencing technology.
While many \emph{de novo} assembly software packages are available, comparatively little attention has been paid to assisting the user with the assembly.
This paper addresses the practical aspects of \emph{de novo} assembly by introducing new ways to perform quality assessment on a collection of DNA sequence reads.
The software implementation calculates per-base error rates, paired-end fragment size histograms and coverage metrics in the absence of a reference genome.
Additionally, the software will estimate characteristics of the sequenced genome, such as repeat content and heterozygosity, that are key determinants of assembly difficulty.
The software described is freely available and open source under the GNU Public License.

\section{Introduction}

The availability of inexpensive DNA sequence data has led to a vast increase in the number of genome projects.
For example, the Genome10K project \cite{Genome10KCommunityofScientists2009Genome} aims to sequence 10,000 vertebrate genomes in the upcoming years.
Despite the advances in the production of DNA sequence data, performing \emph{de novo} assembly remains a significant challenge.
This challenge was highlighted by the recent Assemblathon2 project \cite{Bradnam2013Assemblathon}.
In this competition sequence data was obtained for three vertebrate genomes.
Twenty-one teams contributed assemblies of the three genomes, producing 43 assemblies in total.
The quality of the assemblies were highly variable both between submissions for the same genome and within individual software packages across the three species.
In my view, this variability stems from the practical difficulty of designing an assembly strategy (for instance, selecting software and its parameters) when little is known about the structure of the underlying genome and the quality of the available data.
This paper aims to address this uncertainty.

Most current genome assemblers are based on constructing a graph representing the relationship between sequence reads or their subsequences.
The sequence of the underlying genome is modelled as a walk (or a set of walks) through the graph.
The properties of the sequenced genome and quality of the input data is reflected by the structure of the graph; repeats, sequence variation (in a diploid or polyploid genome) and sequencing errors cause branches in the graph.
These branches increase the difficulty of the assembly by obscuring the true walks that represent the sequence of the genome with many false alternatives.
Below, we will show how we can estimate the individual contribution of sequence variants, repeats and sequencing errors to the branching structure of an assembly graph and we will discuss how the branching structure impacts assembly difficulty.
Additionally, we will develop methods to perform rich quality assessment without a reference genome, complementing previously developed approaches \cite{FastQC, Schroder2010ReferenceFree, Wang2012Estimation, Keegan2012PlatformIndependent} by estimating sequence coverage, per-base error rates, insert size distributions and providing a visual method to assess coverage biases due to sequence composition \cite{Dohm2008Substantial, Ross2013Characterizing}.

Our software is open source under the GNU Public License (version 3) and freely available online (\url{https://github.com/jts/sga}).
The implementation uses the FM-index data structure, which allows queries to be performed over a large text collection while limiting memory usage.
This framework allows our analysis pipeline to be run on 170 Gbp of human genome data in under 24 hours using 56 GB of memory on a single multi-core computer.
The output of our software is a PDF report that allows the properties of the genome and data quality to be visually explored.
By providing more information to the user at the start of an assembly project, this software will help increase awareness of the factors that make a given assembly easy or difficult, assist in the selection of software and parameters and help to troubleshoot an assembly if it runs into problems.

\section{Results}

\subsection{Input data}

In the following sections, we demonstrate the output of our program using freely available data from genomes of varying difficulty.
The selected data sets and their accessions are:

\begin{itemize}
\item \emph{Saccharomyces cerevisiae} (ERR049929)
\item \emph{Melopsittacus undulatus}, a budgerigar from Assemblathon2 (ERR244146)
\item \emph{Maylandia zebra}, a Lake Malawi Cichlid from Assemblathon2 (SRX033046)
\item \emph{Boa constrictor constrictor}, a snake from Assemblathon2 (ERR234359-ERR234374)
\item \emph{Crassostrea gigas}, a Pacific oyster (SRR322874-SRR322877)
\item \emph{Homo sapiens}, a human genome (ERR091571-ERR091574)
\end{itemize}

For simplicity and consistency with the Assemblathon2 paper we will refer to these data sets as `yeast', `bird', `fish' `snake', `oyster' and `human'. 
The yeast genome was selected to provide an example of an uncomplicated genome that is typically straightforward to assemble.
In contrast, the oyster genome is highly heterozygous and repeat-rich.
The genome was recently sequenced using a fosmid-pooling strategy after whole genome assembly failed to produce satisfactory results \cite{Zhang2012Oyster}.
The human and Assemblathon2 data sets represent a range of large eukaryotic genomes of varying heterozygosity and repeat content.
Multiple high-coverage sequencing libraries are available for the human and Assemblathon2 samples.
For each genome a single library was selected for analysis.
For the oyster data, all three short-insert libraries are used for the inbred sample to provide adequete coverage to infer the properties of the genome.
The yeast data set was downsampled from 500X coverage to 40X to be consistent with the other data sets.
We first describe our estimates of genome characteristics, followed by our data quality metrics.

\subsection{Exploring heterozygosity}

\begin{figure}[p]
\center
\includegraphics[width=\textwidth]{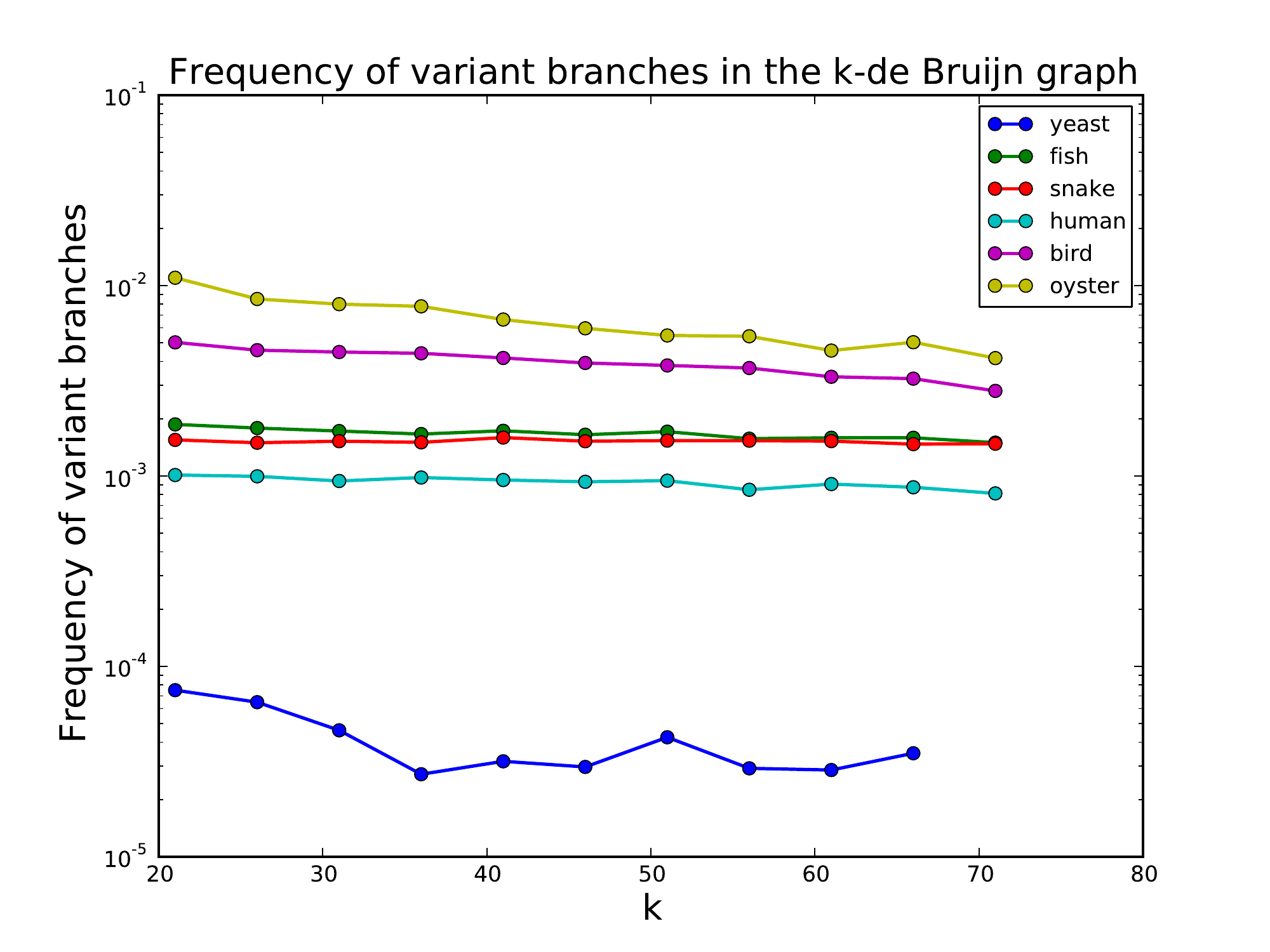}
\caption{The estimated variation branch rate for each genome as a function of $k$}
\label{fig_variant_branches}
\end{figure}

Allelic differences in a diploid or polyploid genome generate branches in an assembly graph with a characteristic structure known as ``bubbles'' \cite{Zerbino2008Velvet}.
Most graph-based assemblers have functions to search for these structures in the graph and remove them.
While these algorithms are typically effective at removing isolated allelic differences, high density variation can make assembly challenging \cite{Donmez2011Hapsembler, ThePotatoGenomeSequencingConsortium2011Genome, Zhang2012Oyster}.
To quantify and visualize the effect of sequence variation on the structure of an assembly graph, we developed a method to search for branches in a de Bruijn graph then use a probabilistic classifier to estimate whether the branch was caused by a sequencing error, a sequence variant or a genomic repeat.
The use of this classifier allows us to separate the contributions of the three types of branches to the structure of an assembly graph, giving insight into the structure of the genome.
Figure \ref{fig_variant_branches} depicts the rate of variant branches in a de Bruijn graph as a function of $k$.
Approximately 1 in 1000 vertices in the de Bruijn graph of the human sample has a variant-induced branch, which is consistent with the rate of heterozygous variation found by reference-based analysis of this genome (see section \textbf{Model accuracy}).

Even without any prior knowledge about the six test genomes, it is easy to see from figure \ref{fig_variant_branches} that the oyster genome has the highest density of variant branches, indicating the genome is highly heterozygous.
As observed in \cite{Zhang2012Oyster} this extreme heterozygosity makes assembly significantly challenging.
Of the three assemblathon genomes, the bird genome has the highest heterozygosity while the fish and snake data sets had similar estimated heterozygosity.
The human genome contains the least level of variation within the diploid species.

A low level of branching in the yeast data set is attributed to sequence variation ($<10^{-4}$ branch rate).
As the sequenced yeast was haploid, these likely represent misclassification of systematic sequencing errors or repeats.

\subsection{Exploring genome repeat content}

Genomic repeats also cause branches in the assembly graph.
As repeat branches tend to be difficult to resolve, often requiring long-range paired end data to jump over the repetitive region \cite{Weber1997Human}, the number of repeat-induced branches is a key indicator of assembly difficulty \cite{Kingsford2010Assembly}.

\begin{figure}[p]
\center
\includegraphics[width=\textwidth]{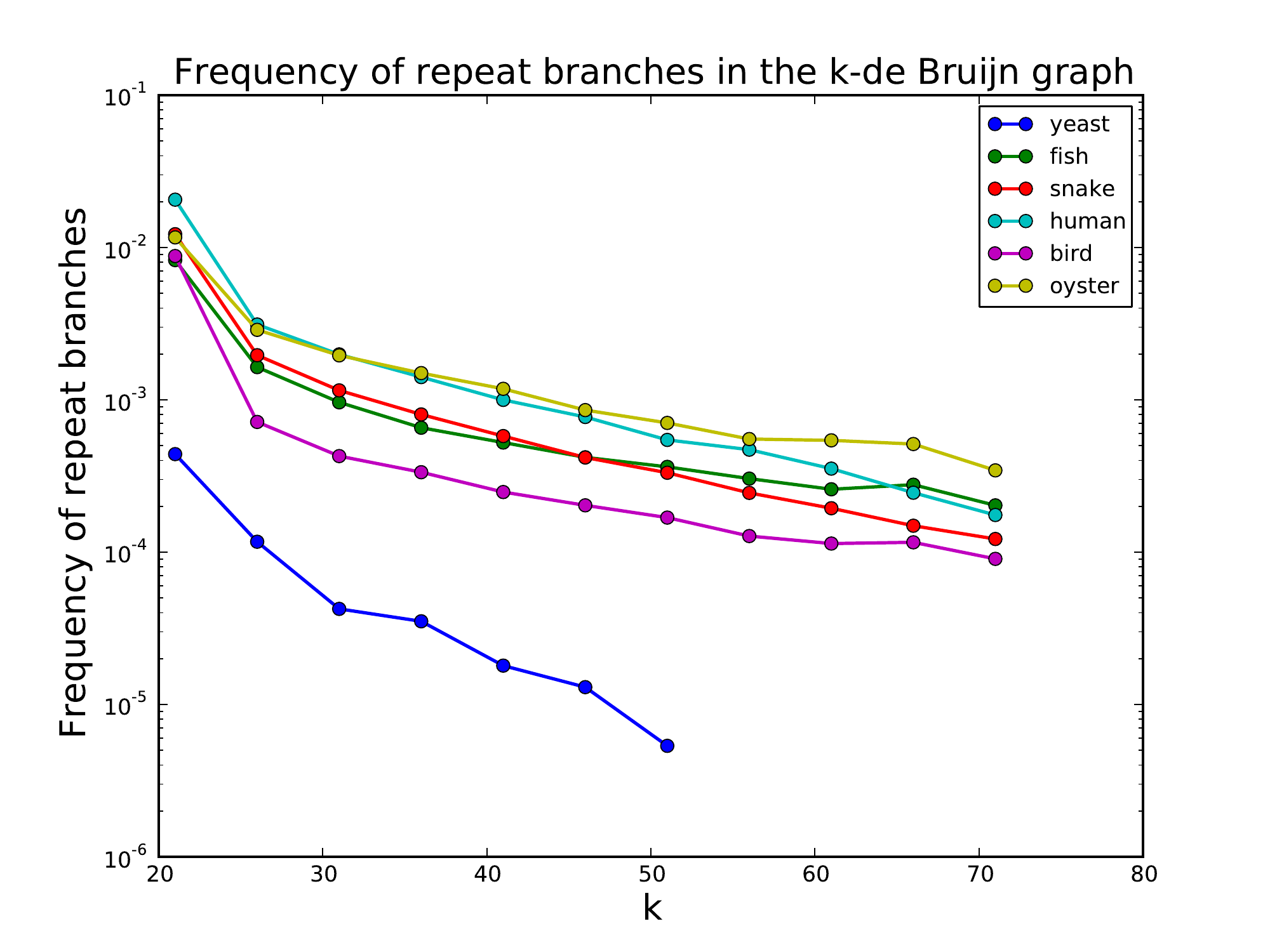}
\caption{The estimated repeat branch rate for each genome as a function of $k$. The yeast data stops at k=51 as the number of repeat branches found falls below the minimum threshold for emitting an estimate.}
\label{fig_repeat_branches}
\end{figure}

We use the output of our classifier to estimate the rate at which repeat-induced branches appear in a de Bruijn graph as a function of $k$ (figure \ref{fig_repeat_branches}).
As expected, the rate of repeat-induced branches clearly decreases as a function of $k$ for all data sets.
The difficulty of assembling the oyster genome is again reflected in the branch analysis.
Despite the genome size being approximately one-fifth the size of the human genome, the oyster genome has a comparable rate of repeat-induced branches.
Likewise, the fish genome is more repetitive than what might be expected from its relatively small genome size.

The yeast genome branches very infrequently due to repeats.
Coupled with the lack of variation shown in the previous section, this suggests that even with small $k$ the de Bruijn graph of the yeast data is relatively uncomplicated and should be straightforward to assemble.

\subsection{Estimating genome size}

The final genome characteristic that we estimate is the size of the genome itself.
Previously, genome size has been estimated from the distribution of $k$-mer counts \cite{Li2010Sequence}.
In the methods section we design a similar method that explicitly corrects for sequencing errors.
Table \ref{table_genome_size} presents a comparison of our genome size estimates to either the reference size or a recent published estimate.

\subsection{Assessing genome coverage}

To facilitate genome assembly, the genome must be sequenced redundantly.
The parameters key to the success of an assembly, particularly the overlap length or $k$-mer size in de Bruijn assembly, are tightly linked to the depth of coverage.
If the parameters to the assembler are too stringent, for instance large $k$ or long overlaps are requested, then the graph may become disconnected due to lack of coverage.
Conversely, if the parameters are too permissive then the graph may contain an unacceptable number of repeat branches.
The parameters are usually chosen (or learned from the data) to maximize stringency subject to available coverage.

\begin{figure}[p]
\center
\includegraphics[width=\textwidth]{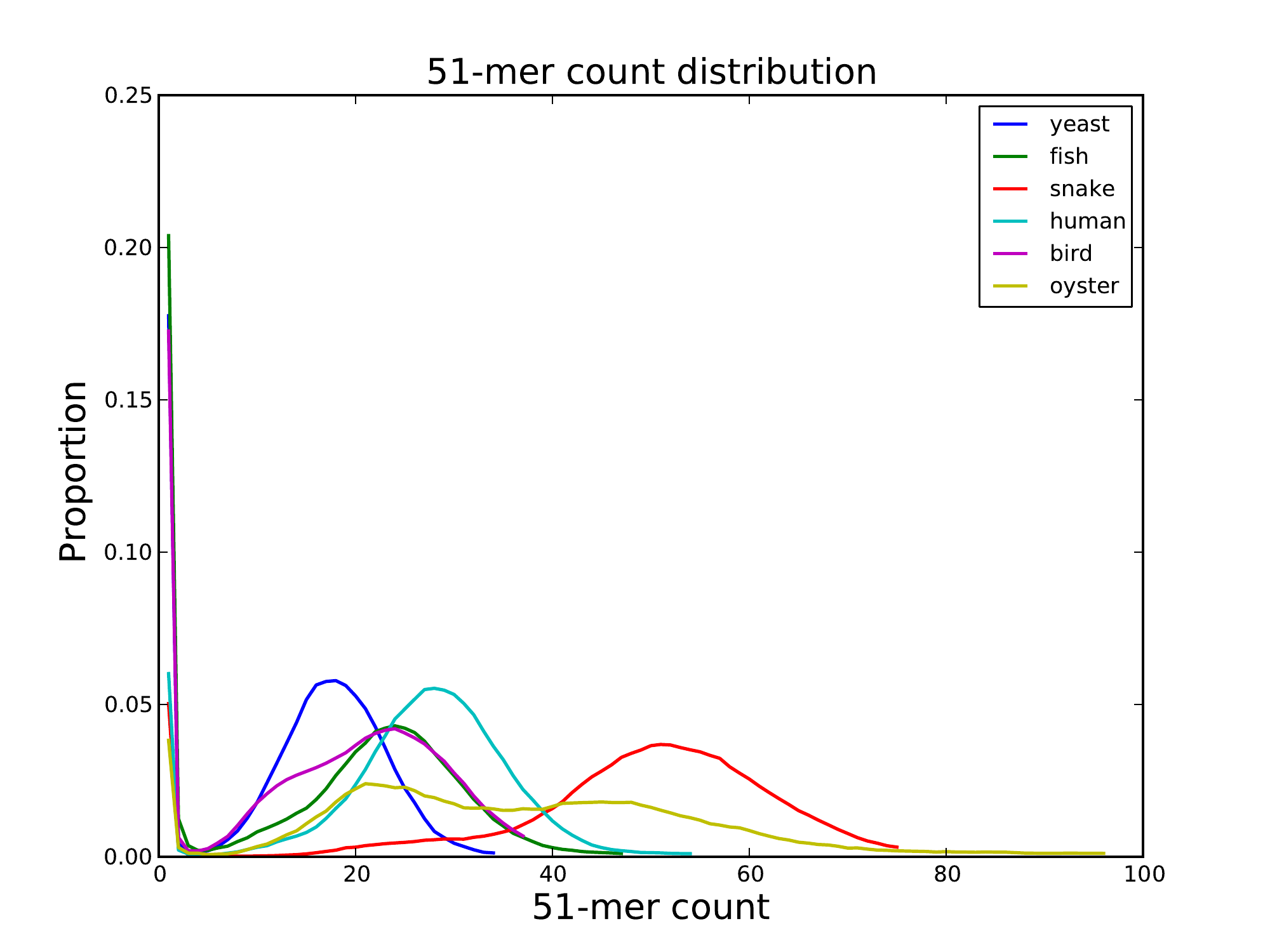}
\caption{A histogram of $51$-mer frequencies for each data set}
\label{fig_count_distribution}
\end{figure}

We have developed multiple methods to assess the coverage of a given data set.
The first method is a histogram of $k$-mer counts for a fixed $k$ (by default $k=51$).
An example is shown in figure \ref{fig_count_distribution}.
On the x-axis are $k$-mer counts and the y-axis is how frequently $k$-mers seen $x$ times occur in the sampled data.
For example, $5-20\%$ percent of $k$-mers are seen only once.
These $k$-mers at low occurrence count typically contain sequencing errors \cite{Pevzner2001Eulerian, Simpson2012Efficient, Kelley2010Quake}.
The remaining $k$-mers, those with higher occurrence count ($>5$ occurrences), are typically error-free and form the substrate of the assembly graph.
Ideally the error-free $k$-mers are well separated from $k$-mers containing errors to allow easy identification and correction of errors.
The snake data is an excellent example of the desired separation, while the yeast data would benefit from more sequencing data or choosing a smaller $k$.

The count distribution also informs our understanding of heterozygosity.
The oyster and bird data, which had the highest estimated heterozygosity by our branch-classification method, have two noticeable peaks in the distribution of error-free $k$-mers.
One peak corresponds to $k$-mers present on both parental haplotypes (at count $46$ for oyster, $24$ for bird) and one peak for $k$-mers covering heterozygotes ($22$ for oyster, $13$ for bird).
The oyster heterozygosity is so high that the peak at count $22$ is the mode of the error-free $51$-mer distribution.

\begin{figure}[p]
\centering
\begin{tabular}{c}
  \subfloat[]{\includegraphics[width=8cm]{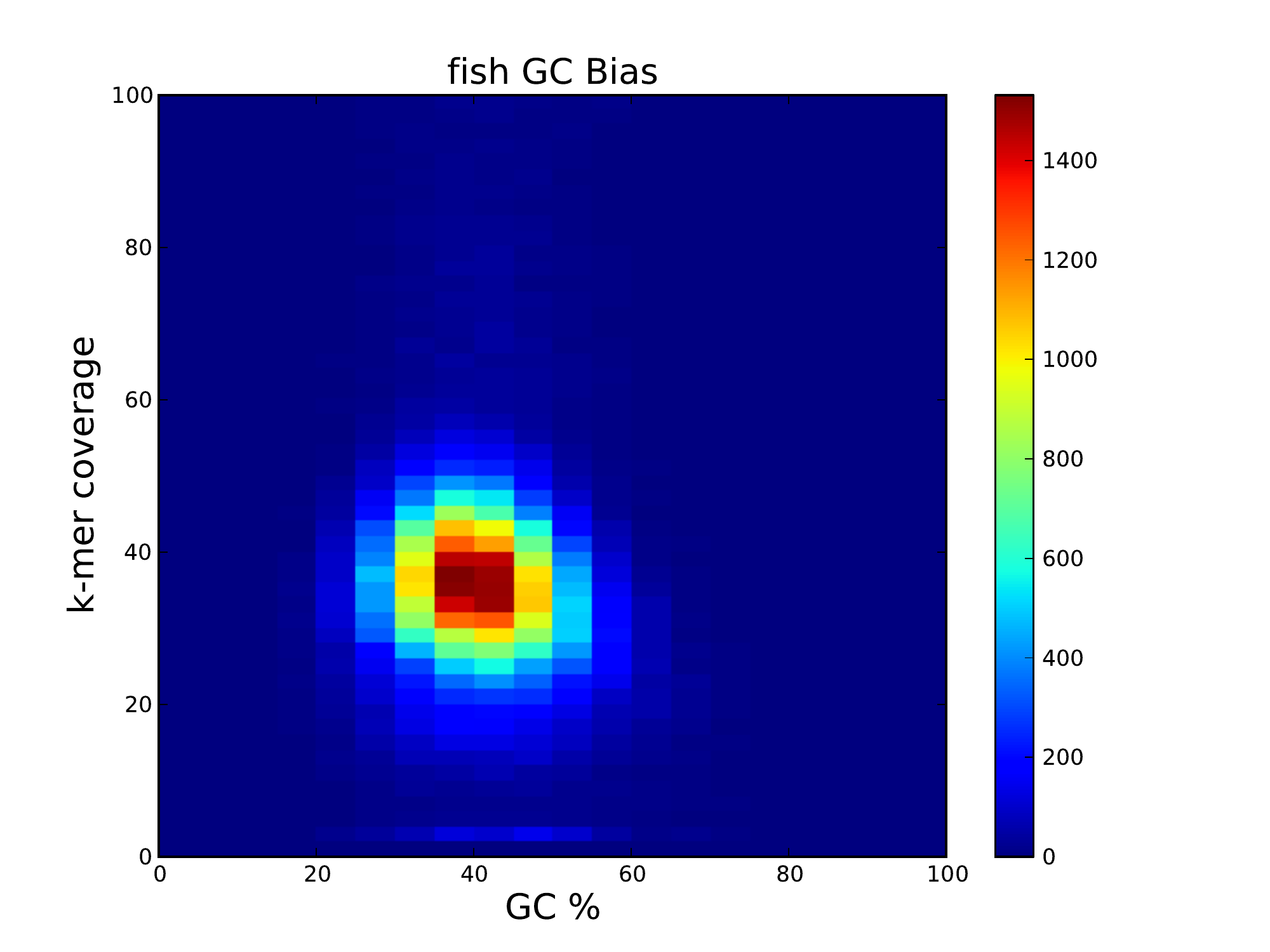}} \\
  \subfloat[]{\includegraphics[width=8cm]{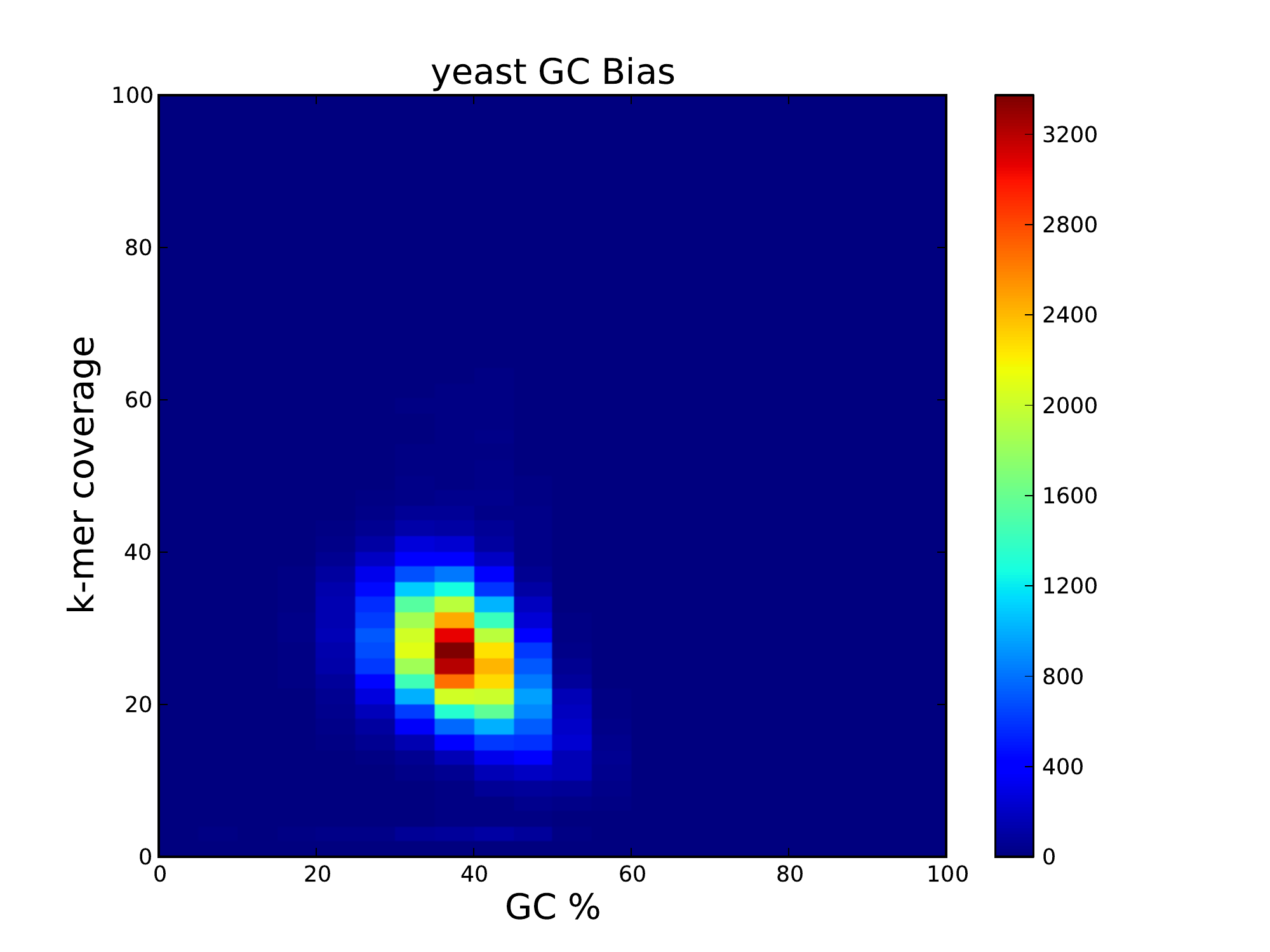}} \\
  \subfloat[]{\includegraphics[width=8cm]{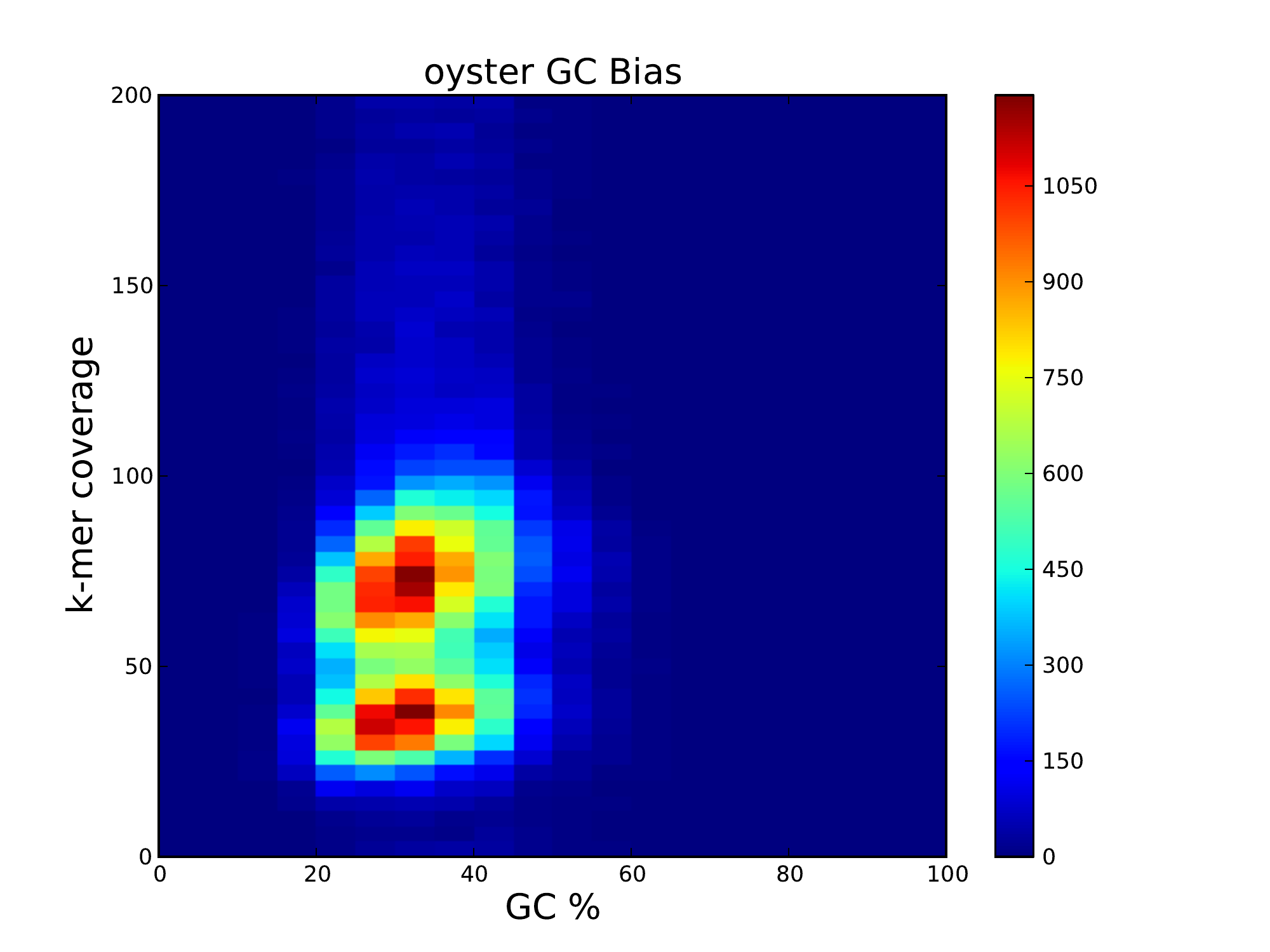}}
\end{tabular}
\caption{Two-dimensional histogram of (GC, count) pairs for $31$-mers for the fish (a), yeast (b) and oyster (c) data sets}
\label{fig_gc_bias}
\end{figure}

Sequence coverage is known to be dependent on the GC content of the sampled fragment \cite{Ross2013Characterizing}.
For extremely biased genomes it can be difficult to cover the entire genome with sequence reads \cite{Kozarewa2009Amplificationfree}.
To visually assess coverage as a function of GC content, we generate a two-dimensional histogram of (GC-content, $k$-mer count) pairs.
If sequence coverage is independent of GC content then the distribution of sequence coverage within each column will be the same.
As an example, the fish data has a relatively uniform coverage profile across the range of GC content (figure \ref{fig_gc_bias}a).
The yeast data is skewed with higher GC sequences having lower coverage on average (figure \ref{fig_gc_bias}b). 
However, the overall coverage is high enough that this mild bias likely does not impact the assembly.
The heteroyzgosity of the oyster data is clearly visibile as two distinct clusters of $k$-mers (figure \ref{fig_gc_bias}c).

\subsection{Simulating contig assembly}

We also designed a method to simulate the output of a de Bruijn assembler to allow the dependency between $k$-mer size and contiguity to be explored.
Recall that for small $k$ the graph will branch more often due to repeats than for large $k$ but for large $k$ we are less likely to sample the complete set of genomic $k$-mers leading to coverage breaks.
Our simulation allows the balance between these factors to be explored by performing walks through a de Bruijn graph mimicking the performance of an assembler that is able to identify and resolve false branches that are caused by errors and bubbles that are caused by variants.
The simulation randomly selects a $k$-mer to seed a new walk through the graph representing an assembled contig.
The simulator extends the contig until a repeat branch in the graph is found (as annotated by our classifier) or extension is stopped due to lack of coverage.

\begin{figure}[p]
\center
\includegraphics[width=\textwidth]{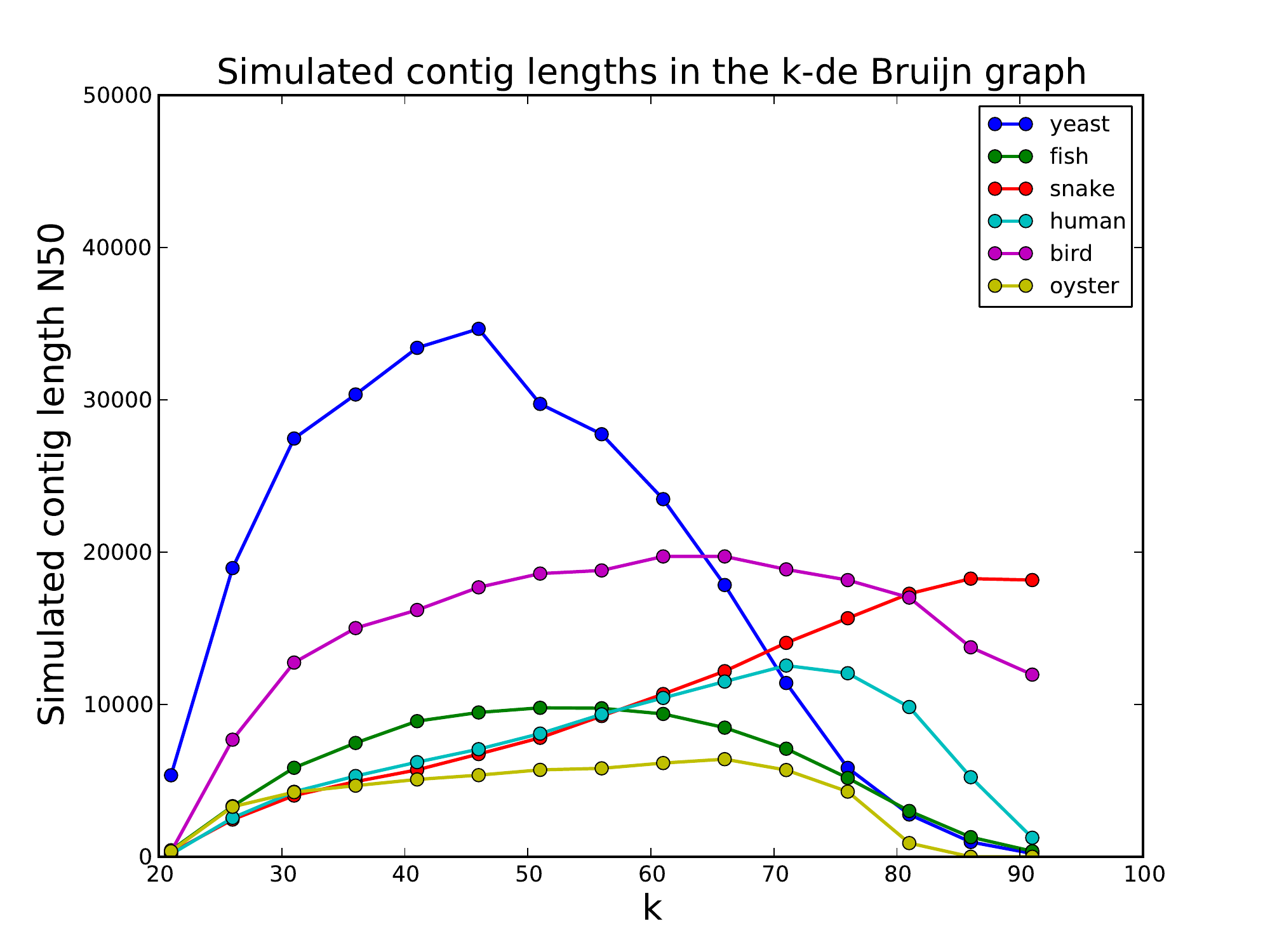}
\caption{The N50 length of simulated contigs for $k$ from 21 to 91, in increments of 5}
\label{fig_simulation}
\end{figure}

The N50 length of simulated contigs as a function of $k$ is shown in figure \ref{fig_simulation}.
For most data sets there is a value of $k$ that maximizes N50 length by striking a balance between ability to resolve short repeats and ensuring the graph is well-connected.
The yeast data is the best example of this with a sharp peak at intermediate $k$.
The snake data is able to support a very large $k$ as the high sequencing depth ensures the graph remains well-connected even for large $k$.
By this assessment, the oyster data is again the most difficult to assemble.

\subsection{Assessing error rates and insert sizes}

\begin{figure}[p]
\center
\includegraphics[width=\textwidth]{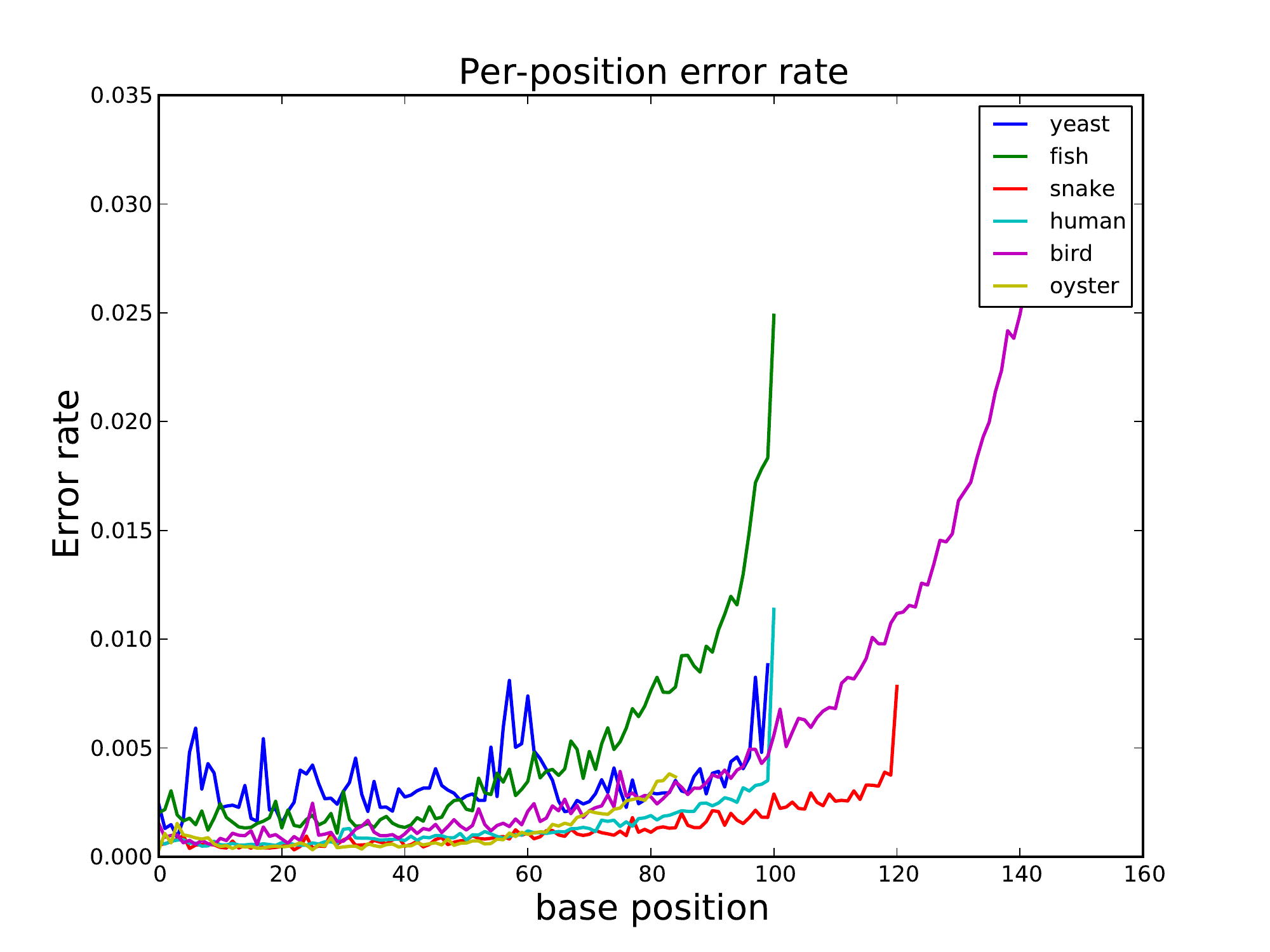}
\caption{The per-base error rate for each data set calculated by read-read overlaps}
\label{fig_error_rate}
\end{figure}

\begin{figure}[p]
\center
\includegraphics[width=\textwidth]{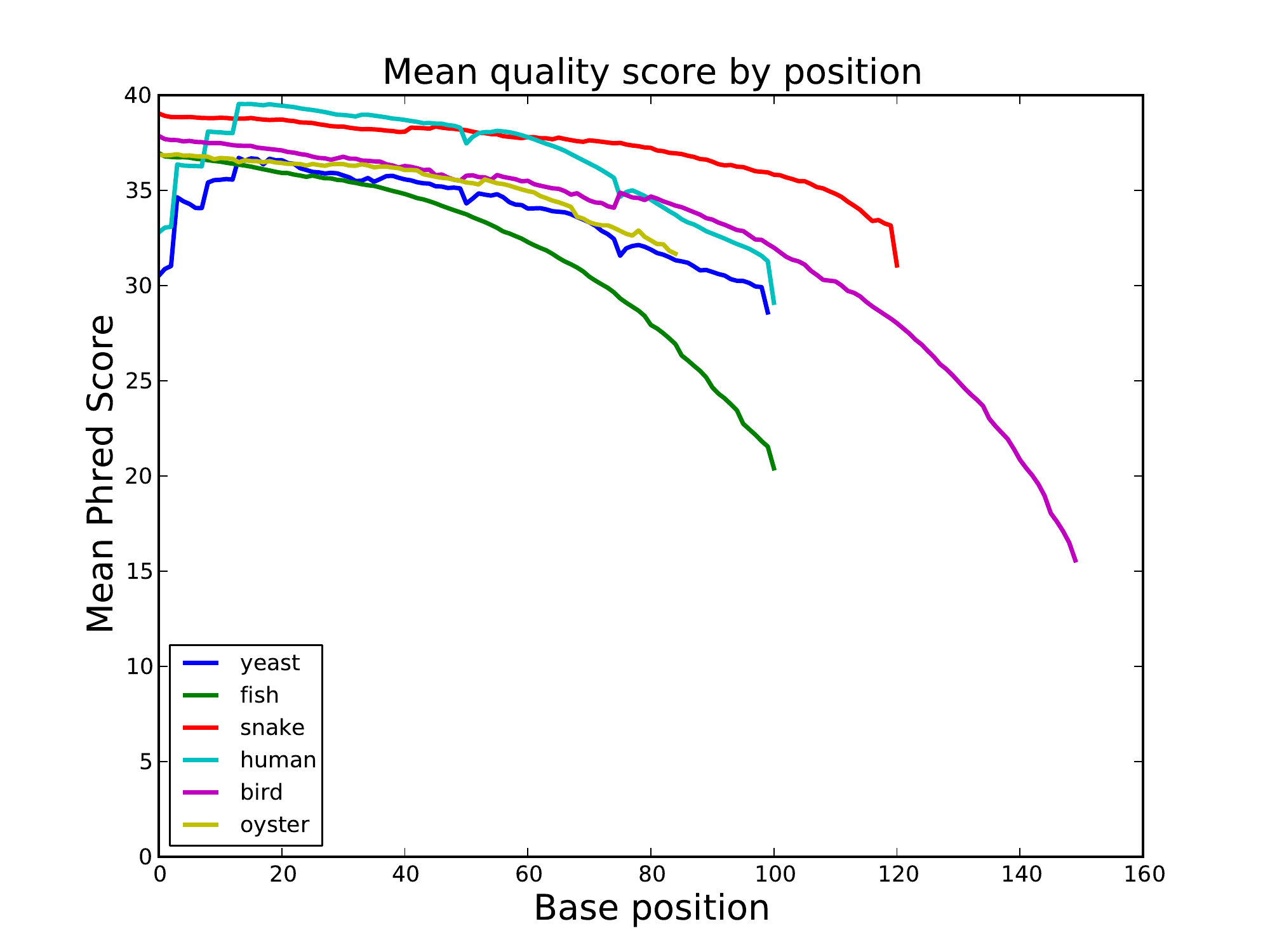}
\caption{The per-base mean quality score for each data set}
\label{fig_mean_quality_score}
\end{figure}

To infer per-base sequencing error rates, we calculate read-read overlaps and compare each read to the consensus sequence of reads it overlaps.
All data sets show the tendency of higher error rate towards the 3' end of the sequence read characteristic of Illumina data \cite{Nakamura2011Sequencespecific} (figure \ref{fig_error_rate}).
Most data sets have less than 1\% error rate across the length of the read.
The distribution of quality scores along the length of the reads shows a similar trend (figure \ref{fig_mean_quality_score}).

\begin{figure}[p]
\center
\includegraphics[width=\textwidth]{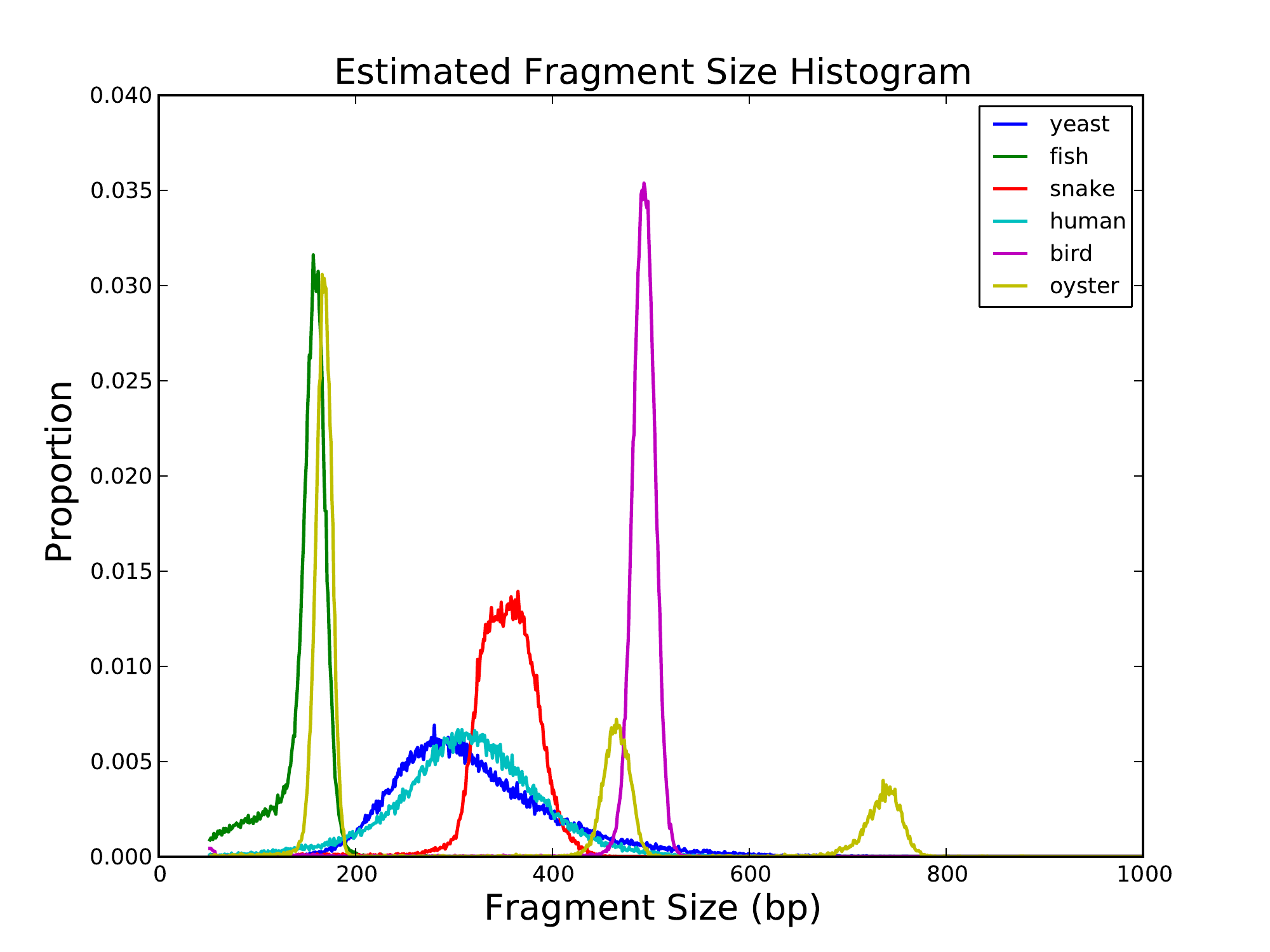}
\caption{The estimated paired-end fragment size for each data set. The oyster data set is a mixture of three libraries.}
\label{fig_fragment_size}
\end{figure}

Paired-end sequence data is commonly used to help span repeats by providing longer-range information.
To help ensure that the sequenced fragments matches the expected size determined by the DNA library preparation, we infer the insert size distribution by performing walks through the assembly graph that begin and end on either end of a read pair (figure \ref{fig_fragment_size}).
In this figure, the oyster data has three modes as it is a mixture of three separate paired-end libraries.

\subsection{Model accuracy}

Finally, we performed a simulation to test the accuracy of our branch classification model.
We performed this assessment by obtaining a diploid reference genome for the human sample (see Methods).
We directly calculated the variant and repeat branch rate from the de Bruijn graph of the diploid reference genome.
We also simulated 40X coverage of this diploid reference and estimated branch rates using the same methodology as the real data.
We expect that the branch rate estimates from the simulated data should match the direct calculations from the reference graph.
The branch rates for the real data should be close to those of the reference graph and simulated data but may differ slightly due to the way the diploid reference genome was prepared.

\begin{figure}[h]
\centering
\begin{tabular}{c}
  \subfloat[]{\includegraphics[width=10cm]{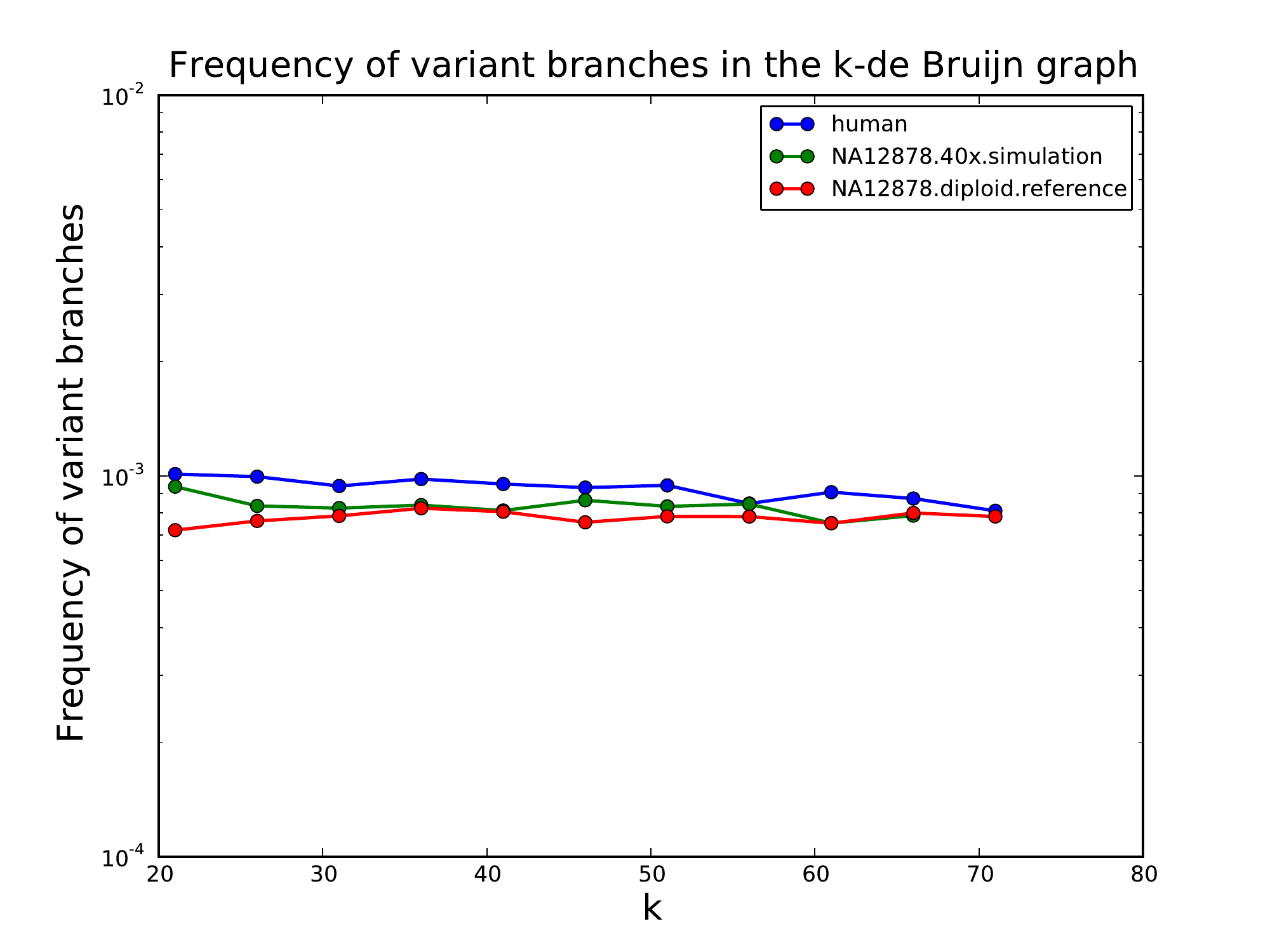}} \\
  \subfloat[]{\includegraphics[width=10cm]{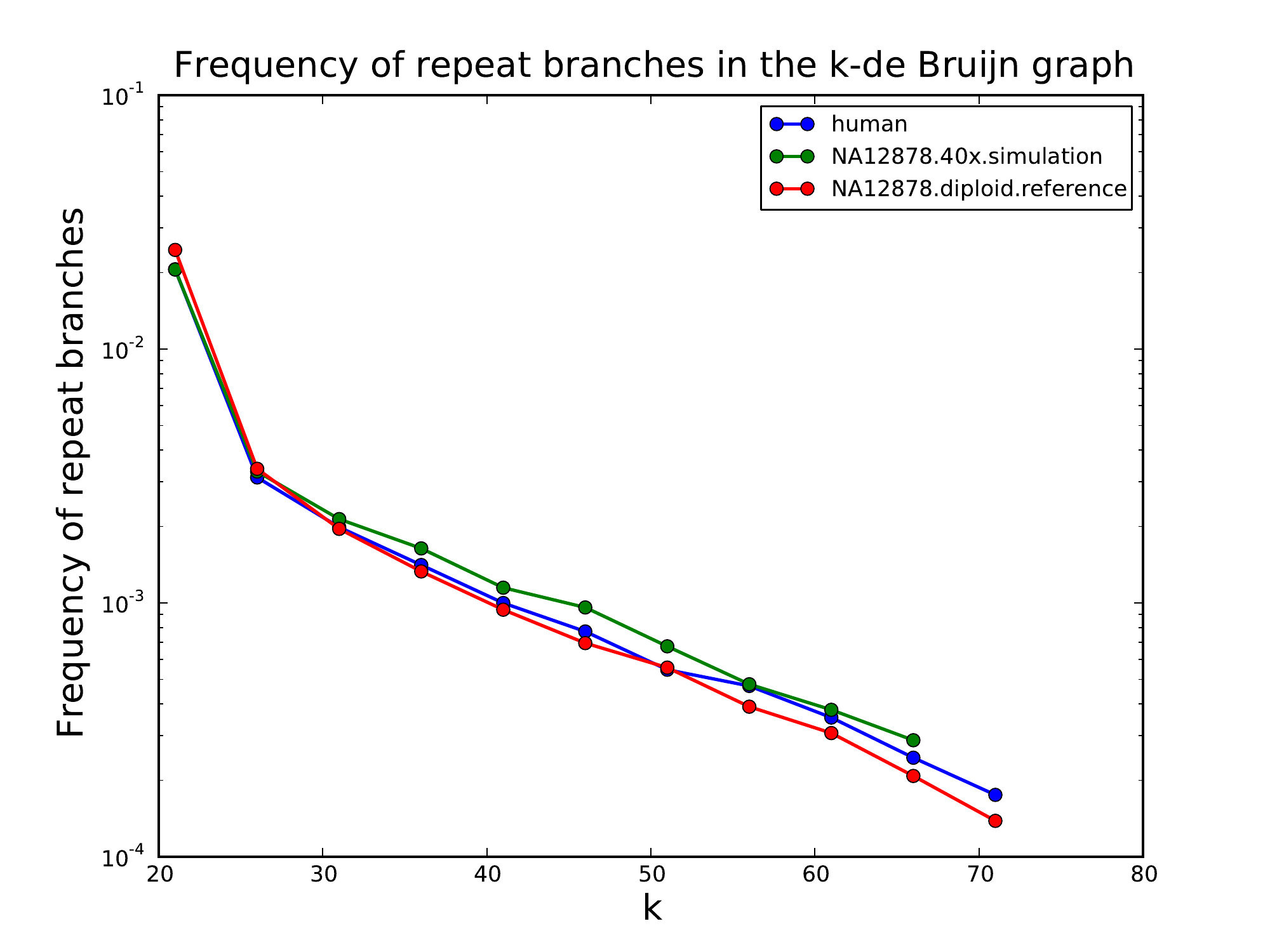}}
\end{tabular}
\caption{Variation (a) and repeat (b) branch rate estimated from real data, simulated data and the diploid reference genome of NA12878}
\label{fig_accuracy_simulation}
\end{figure}

In figure \ref{fig_accuracy_simulation}, the variant and repeat branch rates for the reference graph, simulated reads and real NA12878 data is shown.
The estimated repeat branch rate for the simulated data and real data closely match the repeat rate of the diploid reference genome.
The variant branch rate for the simulated data closely matches the reference calculation, except for very low $k$. 
At low $k$ there is a very high density of repeat branches, which suggests misclassification of repeats may lead to an overestimation of the variant branch rate.
The variant branch rate for the real data set is consistently higher than the simulation and direct reference calculation.
This difference may be due to misclassification of systematic sequencing errors as variants or indicate that an imcomplete variant set was used to construct the diploid reference genome.

\section{Discussion}

While the development of new genome assembly methods continues, comparatively little attention has been paid to assisting the user from a practical standpoint.
Our program, along with tools like VelvetOptimiser \cite{VelvetOptimiser} and KmerGenie \cite{Chikhi2013Informed}, attempts to fill this gap.
The program we developed helps the user perform quality checks on their data while simulatenously assessing the difficulty of the assembly by measuring the branching structure of a de Bruijn graph.
By helping the user better understand their data, our program makes progress towards the goal of making assembly an easier and more consistent process.

This work extends the branch classification model developed by Iqbal et al \cite{Iqbal2012De}.
Most assemblers currently use heurestics to determine whether a branch in the graph is caused by a sequencing error or variant.
The branch classification models could be used in place of these heurestics which may improve assembler performance by adapting the algorithms to the genome structure or data quality.

\section{Methods}

\subsection{Framework}

This work is based on performing queries over a large collection of reads, $S$.
The basic building block of the following methods is simply counting the number of times a particular string occurs in the collection.
For this task, we use the FM-index \cite{Ferragina2000Opportunistic} which allows the number of occurrences of a pattern $P$ in the text $S$ to be counted in time proportional to the length of $P$.
We will use the notation $\algo{count}(P)$ to refer to this procedure.
As our data set consists of DNA and we will often want to know the count of $P$ and its reverse-complement, we define the function 

\begin{equation*}
\algo{countDNA}(P) = \algo{count}(P) + \algo{count}(\algo{reverseComplement}(P))
\end{equation*}

A second building block of our algorithms is sampling a read at random from the FM-index of the read collection $S$.
We adapted the well-known functions to efficiently extract arbitrary substrings of the text from the FM-index \cite{Ferragina2004AlphabetFriendly} to the restricted case of extracting an entire read from $S$.
We will call the procedure to extract read $i$ from the index $\algo{extract}(i)$.
For a read collection with $n$ reads, our sampling procedure simply draws a random number $i$ from $0$ to $n-1$ then runs $\algo{extract}(i)$.

We can also use the FM-index to implicitly represent the structure of a de Bruijn graph.
In Pevzner's original definition of a de Bruijn graph $k$-mer subsequences of the reads are nodes in the graph \cite{Pevzner2001Eulerian}. 
Two nodes $X$ and $Y$ are connected by an edge if some read contains a $(k+1)$-mer that contains $X$ as a prefix and $Y$ as a suffix, or vice-versa.
This condition allows one to formulate the assembly as a tour of the graph that visits each edge at least once.
As we do not require this condition for this work, we adopt the slightly simpler definition of the graph where the vertex set is the set of $k$-mer subsequences and the edges are defined by $k-1$ overlaps between $k$-mers \cite{Simpson2009ABySS, Pell2012Scaling}.
For our purposes, we consider a $k$-mer and its reverse complement to be the same vertex.

This definition of the graph allows us to determine the structure of the graph by simply performing $k$-mer count queries on the FM-index.
Given a vertex sequence $X$, we can use the following procedure to find the neighbors of $X$.
Write $X$ as $X=aZ$ where $Z$ is the $k-1$ suffix of $X$. We can then run $\algo{countDNA}(Zb)$ for $b \in ACGT$.
The $k$-mers with non-zero count represent the \emph{suffix neighbors} of $X$.
The \emph{prefix neighbors} of $X$ can be found similarly.

If a vertex has multiple suffix neighbors, we call it a \emph{suffix branch} (respectively, \emph{prefix branch}).

\subsection{Learning the $k$-mer occurrence distribution}

In the results section ``Assessing genome coverage'' we display the frequency of $k$-mer occurrence counts.
To calculate the data for this plot, we sample $100,000$ reads from the FM-index at random and run $\algo{countDNA}$ on every $k$-mer in every sampled read.
The number of times each count was seen is recorded and the frequency of each count is plotted.

\subsection{Estimating genome size}

Assuming all reads are length $l$ and the reads do not contain sequencing errors, there is a simple relationship between $k$, the number of reads ($n$) and genome size ($G$) \cite{Li2010Sequence}. 
There are $n(l - k + 1)$ $k$-mers in the reads and $G - k  + 1 \approx G$ (as $G >> k$) $k$-mers in the genome. 
The mean number of times a unique genomic $k$-mer appears in the reads is therefore:

\begin{equation*}
\lambda_k = \frac{n(l - k + 1)}{G}
\end{equation*}

If we know $\lambda_k$, which we approximate using the mode of the $k$-mer count histogram, $G$ can easily be calculated.
If the reads contain sequencing errors, this calculation requires modification.
In this case the quantity $n(l - k + 1)$, the total number of $k$-mers in the reads, is a mixture of genomic $k$-mers and artificial $k$-mers containing sequencing errors.
The mode of the $k$-mer count histogram approximates the mean number of genomic $k$-mers, and does not include $k$-mers containing errors.
Therefore the calculation $G = n(l - k + 1) / \lambda_k$ will overestimate $G$ as $n(l - k + 1)$ is inflated by $k$-mers with errors.
To correct for this we estimate the proportion of $k$-mers in the reads that are genomic $k$-mers, $p$.
Our calculation of genome size then becomes:

\begin{equation*}
G = \frac{pn(l - k + 1)}{\lambda_k}
\end{equation*}

We calculate $p$ as follows. 
The probability that a $k$-mer seen $c$ times in the reads is genomic is given by:

\begin{equation*}
P(\ms{genomic}|c) = \frac{P(c|\ms{genomic}) P(\ms{genomic})}{P(c|\ms{error})P(\ms{error}) + P(c|\ms{genomic})P(\ms{genomic})}
\end{equation*}

For the probability of $c$ conditional on having an error or not, we use Poisson distributions that are truncated to account for the fact that we cannot observe $k$-mers with count zero:

\begin{subequations}
\begin{flalign*}
& P(c|\ms{error}) = \algo{Poisson}(c, \lambda_{k}e) z_e \\
& P(c|\ms{genomic}) = \algo{Poisson}(c, \lambda_k) z_g 
\end{flalign*}
\end{subequations}

\noindent
where $e=0.02$ and $z_e=1 / (1 - \algo{Poisson}(0, \lambda_{k}e))$, $z_g=1 / (1 - \algo{Poisson}(0, \lambda_k))$ are the scaling factors for the truncated Poisson.
We set the priors $P(\ms{genomic}) = 0.9$ and $P(\ms{error}) = 0.1$.

Letting $N_c$ be the number of sampled $k$-mers with count $c$:

\begin{equation*}
p = \frac{\sum_{c=1} P(\ms{genomic}|c) N_c}{\sum_{c=1} N_c}
\end{equation*}

As we assume the $k$-mer counts follow a Poisson distribution the mode of the distribution provides a reliable approximation of $\lambda_k$.
However, when selecting the mode from the $k$-mer count histogram some care is needed.
For data with an extremely high error rate, $k$-mers seen a single time ($c=1$) may be the most frequent.
To avoid selecting the error mode, we select the mode after the first local minimum of the distribution.
For highly heterozygous data, like the oyster data, the mode of the distribution may correspond to heterozygous $k$-mers.
To account for this we explicitly check for a secondary peak.
Let $m$ be the mode of the distribution, which has height $N_m$.
If there is a second peak at $2m$ with height $N_{2m} \geq N_m/2$ we use $2m$ as our approximation of $\lambda_k$ instead of $m$.

In principle, it is preferable to model the count distribution as a mixture of negative binomial distributions to explicitly model the genomic copy number of each $k$-mer and overdispersion of the count data.
As the mixture proportions and parameters to the negative binomial are more complicated to fit, we opt for the simpler model here which only relies on finding the mode of the distribution.

We use $k=31$ to perform our genome size estimates.
To estimate the $31$-mer count histogram, $20,000$ reads are sampled using the same method as the previous section.

\subsection{Branch classification}

To quantify the rate of branching in a de Bruijn graph that can attributed to sequencing errors, sequence variation and repeats we designed an algorithm to discover and classify branches in the graph.
We assume that the sequenced genome is diploid.
We start by sampling a read from the FM-index then iterating over all $k$-mers in the read.
Let $k_i$ be the current $k$-mer and $c_i = \algo{countDNA}(k_i)$.
To calculate branch rate estimates, we only want to classify $k$-mers that are expected to occur a single time on each parental chromosome.
We will call $k_i$ a \emph{homozygous k-mer} if $k_i$ occurs a single time on each parental chromosome.
We conservatively estimate the probability that $k_i$ is a homozygous $k$-mer using a simple Poisson model:

\begin{subequations}
\begin{flalign*}
P(\ms{homozygous}|c_i) = \frac{P(c_i|\ms{homozygous})P(\ms{homozygous})}{P(c_i)} \\
\end{flalign*}
\end{subequations}

\noindent
where: 

\begin{equation}
\begin{split}
P(c_i) = &\quad P(c_i|\ms{homozygous})P(\ms{homozygous}) \\ 
 &\quad + P(c_i|\ms{heterozygous})P(\ms{heterozygous}) \\
 &\quad + P(c_i|\ms{repetitive})P(\ms{repetitive})
\end{split}
\end{equation}

$P(c_i|\ms{homozygous})$, $P(c_i|\ms{heterozygous})$ and $P(c_i|\ms{repetitive})$ are Poisson distributions with parameters $\lambda_k$, $\lambda_k/2$ and $2\lambda_k$ respectively, where $\lambda_k$ is estimated from the $k$-mer count distribution as described above.
We use $1/3$ as the prior probability for each state.

If $P(\ms{homozygous}|c_i) < 0.90$ we discard the $k$-mer otherwise we find the suffix neighbors of $k_i$ as described above.
To minimize the impact of systematic errors \cite{Guo2012Effect}, we require that a neighboring $k$-mer is seen on both sequencing strands to be a valid edge in the graph.
If $k_i$ has more than one suffix neighbor, we attempt to classify the branch.
Let $k_a$ and $k_b$ be the two highest-coverage neighbors of $k_i$ with counts $c_a$ and $c_b$.
We set $k_a$ to be the higher coverage neighbor ($c_a \geq c_b$).

Our classifier is a modified version of the probabilistic model designed by Iqbal et al. \cite{Iqbal2012De}.
Initially we modelled the total coverage of the branch, $t = c_a + c_b$, using a Poisson distribution with mean $\lambda_k$ under the variant and error models and $r\lambda_k$ in the repeat model (for $r \geq 2$, representing the repeat copy number).
This model tended to misclassify repeats as variants in the case when $k_i$ is from a low-coverage region of the genome, as $t$ is correlated to $c_i$ and therefore undersampling $k_i$ biased $t$ to be smaller than expected.
To account for this, we define a new variable without the dependency on $c_i$.
Let $c_{ia}$ (respectively $c_{ib}$) be the number of reads that contain both $k_i$ and $k_a$ ($k_i$ and $k_b$).
We define $d = c_a + c_b - c_{ia} - c_{ib}$.
Intuitively, $d$ is the number of reads that contains $k_a$ or $k_b$ but not $k_i$.
Under the variant and error model this is only possible when $k_a$ or $k_b$ is the first $k$-mer of a read or if there is a sequencing error in the first base of $k_i$.
Both of these cases are relatively rare so $d$ is expected to be very small under the variant and error model.
Under the repeat model $k_a$ and $k_b$ appear in more genomic locations than $k_i$.
This gives more opportunities to sample $k$-mers covering $k_a$ and $k_b$ so we expect $d$ to be relatively large.

We use the following distributions for $d$, conditional on the classification:

\begin{subequations}
\begin{flalign*}
& P(d|\ms{error}) = \algo{Poisson}(d, \lambda_r + \lambda_e) \\
& P(d|\ms{variant}) = \algo{Poisson}(d, \lambda_r + \lambda_e) \\
& P(d|\ms{repeat}) = \sum_{r=2} \algo{Poisson}(d, (r - 1) \lambda_k)P(r)
\end{flalign*}
\end{subequations}

\noindent
where $\lambda_r=n/G$ is the density of read starting positions along the genome and $G$ is the estimated genome size.
$\lambda_e$ is the mean number of occurrences of a $k$-mer that has a sequencing error, which we set to be $e\lambda_k$ as in our genome size estimate.
In the repeat model, we sum over integral copy numbers starting from 2 copies.
For this calculation the probability of having $r$ copies of a $k$-mer, $P(r)$, is given by a geometric distribution with parameter $\mu=0.8$, scaled to account for the fact that we do not allow $r=1$ in the repeat state.

The second source of information is the coverage \emph{balance} between $k_a$ and $k_b$.
If $k_a$ and $k_b$ represent a variant, we expect each $k$-mer to be equally well represented.
If the branch is due to an error we expect most reads to support the higher coverage neighbor, $k_a$.
We model coverage balance with the following distributions:

\begin{subequations}
\begin{flalign*}
& P(c_a, c_b|\ms{error}) = \algo{BetaBinomial}(c_a, c_a + c_b, 50, 1) \\
& P(c_a, c_b|\ms{variant}) = \algo{Binomial}(c_a, c_a + c_b, 0.5) \\
& P(c_a, c_b|\ms{repeat}) = \algo{BetaBinomial}(c_a, c_a + c_b, 5, 1) \\
\end{flalign*}
\end{subequations}

Here, $\algo{BetaBinomial}(k, n, \alpha, \beta)$ is the probability mass function of the Beta-Binomial distribution parameterized by $\alpha$ and $\beta$ and $\algo{Binomial}(k, n, p)$ is the probability mass function of the Binomial distribution parameterized by $p$.
The $\alpha$ and $\beta$ parameters of the $\algo{BetaBinomial}$ under the repeat model are chosen to reflect our uncertainty of the genomic copy number configuration of $k_a$ and $k_b$.

Assuming indepedence of $d$ and $c_a,c_b$, we calculate the posterior probability of each classification for each branch encountered.
We use $1/3$ as the prior probability of each classification when calculating the posterior.
Our output is the expected number of branches of each type ($N_e, N_v, N_r$) and the expected number of homozygous $k$-mers that were checked for a branch ($N_h$).
Initially set to zero, these expectations are updated as follows:

\begin{subequations}
\begin{flalign*}
& N_e \leftarrow N_e + P(\ms{homozygous} | k_i)P(\ms{error} | c_a, c_b, d) \\
& N_v \leftarrow N_v + P(\ms{homozygous} | k_i)P(\ms{variant} | c_a, c_b, d) \\
& N_r \leftarrow N_r + P(\ms{homozygous} | k_i)P(\ms{repeat} | c_a, c_b, d) \\
& N_h \leftarrow N_h + P(\ms{homozygous} | k_i) \\
\end{flalign*}
\end{subequations}

\noindent
$N_e$, $N_v$ and $N_r$ are only updated when a suffix branch in the graph was found.
$N_h$ is updated for every homozygous $k$-mer processed.

We perform this classification on every $k$-mer in $50,000$ randomly sampled reads for $k$ $21$ to $71$ in increments of $5$.
For the output plots in the report, the branch rates are calculated as $N_r / N_h$ and $N_v / N_h$.
If the expected number of branches for a classification is less than $2$, no point is plotted for that value of $k$.

This model has limited power to distinguish between sequence errors and variants when $\lambda_k$ is small.
Additionally, if $\lambda_k$ is too small we will simply not observe variant branches in the graph due to both alleles not being represented in the sequence data.
For this reason, we do not output classifications when $\lambda_k < 10$.

\subsection{Estimating per-base error rates}

To estimate per-base error rates we compute read-read overlaps that are seeded by short exact matches.
We begin by sampling a read $R$ from the FM-index and computing the set of reads that share a $31$-mer with $R$.
For each read sharing a $31$-mer with $R$, we compute an overlap between the read and $R$.
To avoid spurious matches between repetitive sequence we require the overlap is at least $50$bp in length and the percent identity is at least $95\%$.
We construct a multiple alignment using $R$ and the pairwise overlaps for reads meeting this threshold.
We then compute a consensus sequence for each column of the multiple alignment.
A base call $R[i] = b$ is considered to be incorrect if $b$ does not match the consensus base, at least $3$ reads support the consensus base and fewer than $4$ reads support base call $b$.
If $b$ is considered to be incorrect we update the count for mismatches at read position $i$.
For this analysis $100,000$ reads are sampled.
To avoid excessively long computation time for repetitive reads, we skip $31$-mers that are seen more than $200$ times in the reads when computing the candidate overlap set.

\subsection{Calculating per-base quality scores}

The average quality score per base is calculated by sampling every 20th read in the input FASTQ file, up to a maximum of $10,000,000$ reads.

\subsection{GC vs coverage plots}

The plots in figure \ref{fig_gc_bias} are two-dimensional histograms of (GC, count) pairs.
To calculate the input data, we sample $100,000$ reads from FM-index and run $\algo{countDNA}$ on the first $31$-mer of the read.
If the count is one, we reject the read as the first $31$-mer likely represents an error.
Otherwise, we calculate the proportion of GC bases in the entire read and emit a (GC, count) pair.
These pairs are input into the $\algo{histogram2d}$ function of the numpy python library (\url{http://www.numpy.org/}).

\subsection{Estimating the fragment size histogram}

We estimate the fragment size histogram by finding walks through a de Bruijn graph.
We begin by sampling a read pair, $X$ and $Y$, from the FM-index.
Starting from the first 51-mer of $X$, we perform a greedy search of the 51-mer de Bruijn graph by choosing the highest coverage branch as the next vertex in the search.
The search stops when the first 51-mer of $Y$ is found, there are no possible extensions or 1500 iterations have passed.
If a complete walk from $X$ to $Y$ is found, the length of the walk in nucleotides is emitted as the fragment size.
If sequence coverage is low, this method of estimating the fragment size histogram may be biased towards shorter fragments, as it is more likely that a walk representing a long fragment is broken by lack of coverage.
For this analysis, $100,000$ read pairs are sampled.

\subsection{Simulating de Bruijn assembly}

Our simulated de Bruijn graph assembly performs a walk through the graph until a repeat or lack of coverage terminates extension.
We begin by sampling a read at random and calculating the probability that the first $k$-mer of the read is a homozygous $k$-mer as in the branch classification method.
If the probability is less than $0.50$, we discard this read and start again.
Otherwise we begin a new contig starting from the first $k$-mer of the read.

Let $X$ be the current $k$-mer of the contig.
We check $X$ for a branch as in our branch classification method.
If $X$ does not have a branch, or has a branch that is classified as an error or variant, we iterate from the highest-coverage neighbor.
If $X$ does not have a neighbor, or has a repeat branch, we terminate extension of the contig.
This procedure occurs for both the suffix neighbors of the initial $k$-mer and the prefix neighbors.
Once the extension has terminated in both directions the number of $k$-mers visited is output.

To avoid excessively long computation time we cap the maximum walk length at $50,000$ and stop extension if a particular $k$-mer is visited twice.
We also do not allow a given walk to be used multiple times by recording all $k$-mers visited in a bloom filter.
Starting $k$-mers that are present in the bloom filter are skipped.
We perform $20,000$ walks for each $k$ from $21$ to $91$ in increments of $5$

\subsection{Branch classification accuracy assessment}

To assess the accuracy of our branch classifier, a diploid reference genome for NA12878 was downloaded (\url{http://sv.gersteinlab.org/NA12878_diploid/NA12878_diploid_dec16.2012.zip}).
This reference genome was constructed from SNP and indel calls for NA12878 phased onto parental chromosomes \cite{Rozowsky2011AlleleSeq}.
The reference genome was processed to change uncalled bases to a random base.
From this diploid reference genome, we simulated sequence reads using DWGSIM (\url{https://github.com/nh13/DWGSIM}) with the following command:

\texttt{dwgsim -C 20 -r 0.0 -1 100 -2 100 -e 0.0001-0.005 -E 0.0001-0.005 -y 0 -d 300 -s 30 NA12878.diploid.fa prefix}

The simulated data was processed using the same pipeline as the real datasets.
Additionally, we calculated the variant and repeat branch rate from the de Bruijn graph of the diploid reference.
The method is similar to section ``Branch classification'' above, except in the place of the probabilistic model we use simple counts to classify the structures in the graph.
If a sampled $k$-mer has a count of exactly $2$ (one copy on each parental chromosome), we consider it to be a homozygous $k$-mer.
Branches are detected by finding homozygous $k$-mers that have multiple neighbors.
When such a branch is found, we classified the branch as a variant if each neighbor had count $1$ in the diploid reference, otherwise we called the branch a repeat.

\subsection{Computations}

The program to calculate the genome characteristics and qc metrics is implemented as a module of the SGA assembler.
This program writes the results to a JSON file, which is read by a python script to generate the PDF report.
The computations performed in this paper are fully reproducible by downloading and running the following Makefile:

\url{https://github.com/jts/preqc-paper/tree/master/bin/generate_data.make}

The Makefile will download the input data from public repositories, run SGA, then generate the final reports.
Version 0.10.10 of SGA was used to generate the data and figures for this paper.
The JSON-formatted results are availble at:

\url{ftp://ftp.sanger.ac.uk/pub/js18/preqc-paper}

The computation time for the human data, the largest set used in the paper, was 13 hrs (elapsed time) to download the data, 18 hours to build the FM-index and 4 hours to calculate the metrics. The memory highwater mark was 56GB during construction of the FM-index.

\section{Acknowledgements}

The author thanks Richard Durbin, Milan Malinsky, Malcolm Hinsley and Daniel Hughes for early feedback on this work.
Leopold Parts, Zamin Iqbal and Shaun Jackman provided valuable comments on a draft of the manuscript.
The use of the oyster data was suggested by Manoj Samanta.
The project was motivated in part by online discussion of the Assemblathon2 paper centered on a blog post by Titus Brown.

\bibliographystyle{bmc_article}
\bibliography{jts}

\newpage
\section*{Figures}

\clearpage

\clearpage

\clearpage

\clearpage

\clearpage

\clearpage

\clearpage

\clearpage

\clearpage

\section*{Tables}

\begin{table}[p]
\begin{center}
\begin{tabular}{| c | c | c | }
\hline
Genome & Reference-Free Estimate & Published size\\
\hline
yeast & 13 Mbp & 12 Mbp \cite{Goffeau1996Life}\\
oyster & 537 Mbp  & 545-637 Mbp \cite{Zhang2012Oyster}\\
fish & 922 Mbp  & 1000 Mbp \cite{Bradnam2013Assemblathon}\\
bird & 1094 Mbp & 1200 Mbp \cite{Bradnam2013Assemblathon}\\
snake & 1408 Mbp & 1600 Mbp \cite{Bradnam2013Assemblathon}\\
human & 2913 Mbp & 3102 Mbp (GRC37)\\
\hline
\end{tabular}
\caption{The genome size estimates from our method compared to previously published estimates}
\label{table_genome_size}
\end{center}
\end{table}

\end{document}